# Evolutionary global optimization posed as a randomly perturbed martingale problem and applied to parameter recovery of chaotic oscillators

Saikat Sarkar[1], Debasish Roy[1]* and Ram Mohan Vasu[2]

[1]*Computational Mechanics Lab, Department of Civil Engineering*

[2]*Tomography Lab, Department of Instrumentation and Applied Physics*

*Indian Institute of Science, Bangalore 560012, India*

*Corresponding author; email: royd@civil.iisc.ernet.in

**Abstract**

A new global stochastic search, guided mainly through derivative-free directional information computable from the sample statistical moments of the design variables within a Monte Carlo setup, is proposed. The search is aided by imparting to a directional update term, which parallels the conventional Gateaux derivative used in a local search for the extrema of smooth cost functionals, additional layers of random perturbations referred to as 'coalescence' and 'scrambling'. A selection scheme, constituting yet another avenue for random perturbation, completes the global search. The direction-driven nature of the search is manifest in the local extremization and coalescence components, which are posed as martingale problems that yield gain-like update terms upon discretization. As anticipated and numerically demonstrated, to a limited extent, against the problem of parameter recovery given the chaotic response histories of a couple of nonlinear oscillators, the proposed method apparently provides for a more rational, more accurate and faster alternative to most available evolutionary schemes, prominently the particle swarm optimization.

**1. Introduction**

The need for global optimization, wherein the aim is generally to determine the global extrema of possibly non-smooth, non-convex cost (or objective) functionals subject to a prescribed set of constraints, is ubiquitous across a broad range of disciplines, from dynamical systems modeling in science and engineering to drug design and delivery. The outcome of such an exercise may, *inter alia*, yield useful parameter information that renders the performance of a system model optimal in a sense made precise by specifying the cost functional. In the context of most practical problems, this functional could be multivariate, multimodal and even non-differentiable, which together precludes applying a gradient-based Newton-step whilst solving the optimization problem. In contrast to finding a local extremal point, attainment of the global extrema of the objective functional, within the domain of definition (search space) of the parameters or the design variables, may be further challenged by a large dimensionality of the search space. The research opportunities created by the ineffectiveness or inapplicability of the Newton search have been the fertile ground for numerous proposals for global optimization, many of which employ stochastic (i.e. random evolutionary) search with heuristic or meta-heuristic origin [1]. Some of such notable schemes include variants of the genetic algorithm (GA) [2], particle swarm optimization (PSO) [3], ant colony optimization [4] to name a few. In the task of finding the global extrema, stochastic search schemes generally score over their gradient-based counterparts [5, 6], even for cases involving sufficiently smooth cost functionals that enable computing well defined directional derivatives. However, as long as the search is only for the nearest local extremum and the objective functional remains differentiable, gradient-based methods offer the benefit of a substantively faster convergence to the nearest extremum owing to the directional information contained in the update equations. To the authors' knowledge, none of the evolutionary global optimization schemes is equipped with an equivalent of such a well defined search direction.

Towards the global search, most evolutionary schemes depend on random scatter applied to the available candidate solutions and the criteria to decide if the new candidates are acceptable. Such randomly drawn candidate solutions are also called 'particles'. A 'greedy' selection for the new particles (i.e. choosing those offering more favorable values of the cost functional), though tempting from the perspective of faster convergence, may encounter the pitfall of getting trapped in local extrema. Most evolutionary global search methods have built-in safeguards against this problem of premature convergence. However, despite the popular appeal of several evolutionary methods of the heuristic/meta-heuristic type [7], the justifying arguments for these schemes often draw sustenance from sociological or biological metaphors [1, 2, 3, 4] that may lack a sound probabilistic basis, even as a random search often forms the cornerstone of the algorithm. Despite this, the wide adoption of these methods owes as much to the algorithmic simplicity as to an efficient global search, which is often accomplished far more effectively than some of the more well-founded stochastic search techniques, e.g. simulated annealing [8], stochastic tunneling etc. [9]. But the absence of a proper probabilistic basis for a (meta-)heuristic scheme may engender a crisis of confidence in its assuredly uniform performance across a broad range of problems and difficulties in enforcing some criterion for performance optimality. This may in turn precipitate a slow convergence to the global optima and also require the end-user to tune a large set parameters for enforcing an appropriate *'exploration-exploitation trade-off'*, a term used to indicate the relative balance of the random scatter (diffusion) of the particles vis-a-vis their selection based on an evaluation of the cost functional. Unfortunately, the ideal values of tuning parameters, on which the performance of the scheme often depend crucially, may change from problem to problem. On the other hand, the concept of random evolution, based on Monte Carlo (MC) sampling of the particles, provides a means to efficient exploration of the search space, albeit at the cost of possibly slower convergence [10] in comparison with a gradient based approach. Dispensing with the notion of derivatives, as in the (meta-)heuristic methods, has also the added advantage of wider applicability. Many such schemes, e.g. the GA that is by far the most popular (though not the most efficient) in the lot, typically have steps like 'crossover', 'mutation', 'selection' etc. While 'crossover' and 'mutation' are essentially ways of bringing in variations in the particles, the 'selection' step is used to assign, to each particle, a 'weight' or 'fitness' value (a measure of importance) determined as the ratio of the realized value of the cost functional, upon substitution of the particle, to the available maximum of the same. The fitness values, used to update the particles via selection of the 'best-fit' individuals for subsequent 'crossover' and 'mutation', may be considered functionally analogous to the derivatives used in gradient based approaches. Parallel to the notion of individual fitness in the GA is that of the likelihood ratio (or weight) assigned to a particle, an MC realization of the state, in a class of nonlinear stochastic filters, e.g. the family of sequential importance sampling filters [11]. Although a weight based approach reduces some measure of misfit between the available best and the rest within a finite ensemble of particles, they bring with them the curse of degeneracy, i.e. all particles but one tend to assume negligibly small weights as the iterations [12] progress.

This problem, known as 'particle collapse' in stochastic filtering, can only be arrested by exponentially increasing the ensemble size (number of particles), a requirement that can hardly be met in practical implementations involving large dimensional states [13]. Optimization schemes that replace the weight-based multiplicative update by an additive term, containing the stochastic equivalent of the directional information in a Newton search, bypass particle degeneracy. PSO, one such meta-heuristic search scheme that is known to be a superior performer to most variants of the GA, utilizes an additive particle update aimed at bridging the mismatch with respect to the available best, both local and global based on the particle evolution histories, whilst still preserving the 'craziness' or scatter in the evolution. Once the desired solution is arrived at, the scatter is however expected to collapse to zero. Unfortunately, none of these methods obtain the directional information in a rigorous or optimal manner, which is perhaps responsible for a painstakingly large number of functional evaluations.

In an attempt at framing a probabilistic setting that incorporates a rigorously derived directional update of the additive type, this article demonstrates that the problem of optimization may be generically posed as a *martingale problem* [14] in the sense of Stroock and Varadhan [15], which must however be randomly perturbed to facilitate the global search. Specifically, the local extremization of a cost functional is posed as a martingale problem realized through the solution of an integro-differential equation, which, following the Freidlin-Wentzell theory [16], is randomly perturbed so that the global extremum is attained as the perturbation vanishes asymptotically. The martingale problem, whose solution obtains a local extremum on the cost functional, involves a so called innovation function, which is viewed as a stochastic process parametered via the iterations and must be driven to a zero-mean martingale [14]. The martingale structure of the innovation essentially implies that, by small perturbations of the argument vector (the design variables), the mean of the computed cost functional does not change over successive iterations and hence the argument vector corresponds to a local extremum. It may be noted that, even though the original extremization problem is posed in a deterministic setting, the cost functional as well as its argument vector are treated as stochastic diffusion processes. In order to realize a zero-mean martingale structure for the innovation, the particles are modified based on a change of measures effected through an additive gain-type update strategy. Thus each candidate from the available population is iteratively guided by an additive correction term so as to locally extremize the given cost functional. The gain-like coefficient, which is a replacement for and a generalization over the Frechet derivative of a smooth cost functional, provides an efficacious directional search without requiring the functional to be differentiable. In order to accomplish the global search, we propose an annealing-type update and a random perturbation strategy (herein represented through the so called operations of 'coalescence', 'scrambling' and 'selection') which together efficiently ensure against a possible trapping of particles in local extrema.

The rest of the paper is organized as follows. Section 2 elaborates on the local optimization posed as a martingale problem leading to an integro-differential equation whose solution is a local optimum. In Section 3, the integro-differential equation is disctretized and weakly solved within an MC setting so as to get around its inherent circularity. Section 4 discusses the proposed random perturbation schemes to arrive at the global optima efficiently without getting stuck in local traps. A couple of pseudo-codes are also included in this section for added clarity in the exposition. In Section 5 we compare the performance of the proposed optimization method with the particle swarm optimization (PSO) in the context of extracting parameters of chaotic oscillators based on some sparse data. Finally, the conclusions of this study are put forth in Section 6.

## 2. Local optimization as a martingale problem

In this section, the functional extremization is posed as a martingale problem, which also includes a generic way of satisfying a given set of constraints. However, before adopting a stochastic framework, a few general remarks on the expected functional features of the new evolutionary optimization scheme would be in order.

- The iterative solution is a random variable (defined on the search space) over every iteration.
- Thus, along the iteration axis, the solution process is considered a stochastic process, whose mean should evolve over iterations to the optimal solution.
- Upon convergence, the mean should be constant and hence iteration-invariant. The random fluctuations about the converged mean is thus a zero-mean stochastic process and may be characterized as noise.

In posing the global optimization problem within a stochastic framework, a complete probability space $(\Omega, \mathcal{F}, P)$ [14] is first adopted, within which the solution to the given optimization problem must exist as an $\mathcal{F}$-measurable random variable. Here $\Omega$, known as the population set (sample space), necessarily includes all possible candidate solutions (particles) from which a set of randomly chosen candidates is evolved along the iteration axis $\tau$. The introduction of $\tau$, as a positive monotonically increasing function in $\mathbb{R}$, is required to qualify the evolution of the solution as a stochastic process. A necessary aspect of the evolution is a random initial scatter provided to the particles so as to search the sample space for solutions that extremize the cost functional whilst satisfying the posed constraints, if any. Since the extremal points, global or local, on the cost functional may not be known *a priori*, the particles are updated iteratively (i.e. along $\tau$) by conditioning on the evolution history of the so called extremal cost process based on the available candidates. In other words, denoting by $\mathcal{N}_\tau := \{\mathcal{G}_\xi\}_{\xi \in [0,\tau]}$ the filtration (evolution history) based on the increasing family of sub $\sigma$-algebras generated by the extremal cost process, defined as a stochastic process such that, for any $\xi \in [0, \tau]$, the mean of the associated random variable is given by the available best cost functional across the particles at the iteration step denoted by $\xi$. In case there are equality constraints, $\mathcal{N}_\tau$ also contains the history for the best realized constraint until $\tau$.

Consider a multivariate multimodal cost functional $f(\mathbf{x}): \mathbb{R}^n \mapsto \mathbb{R}$ which is nonlinear in $\mathbf{x} = \{x^j\}_{j=1}^n \in \mathbb{R}^n$. A point $\mathbf{x}^*$ needs to be found out such that $f(\mathbf{x}^*) \leq f(\mathbf{x}), \forall \mathbf{x} \in \mathbb{R}^n$. The vector variable $\mathbf{x}$ is evolved in $\tau$ as a stochastic process, thus allowing for the parameterization $\mathbf{x}_\tau := \mathbf{x}(\tau)$. Since there may not be any inherent physical dynamics in its evolution, $\mathbf{x}_\tau$ may be given a zero-mean random perturbation in $\Omega$ over every iteration. Specifically, since the number of iterations is a finite integer, one discretizes $\tau$ as $\tau_0 < \tau_1 < ... < \tau_M$ and evolve $\mathbf{x}_\tau$ for every $\tau$ increment. Thus, for the $i^{\text{th}}$ iteration with $\tau \in (\tau_{i-1}, \tau_i]$, $\mathbf{x}_\tau$ may be thought of as being governed by the following stochastic differential equation (SDE):

$$d\mathbf{x}_\tau = d\boldsymbol{\xi}_\tau \tag{2.1}$$

Here $\boldsymbol{\xi}_\tau$ is a zero-mean noise process (e.g. a random walk) with the covariance matrix $\mathbf{gg}^T \in \mathbb{R}^{n \times n}$, where $\mathbf{g} \in \mathbb{R}^{n \times n}$ is the intensity matrix with its $(j,k)^{\text{th}}$ element denoted as $g^{jk}$. The discrete $\tau$-marching map for Eqn. (2.1) may be written as:

$$\mathbf{x}_i = \mathbf{x}_{i-1} + \Delta\boldsymbol{\xi}_i \ , \ \Delta\boldsymbol{\xi}_i := \boldsymbol{\xi}_i - \boldsymbol{\xi}_{i-1} \tag{2.2}$$

where $(\cdot)_i$ denotes $(\cdot)_{\tau_i}$ for notational convenience. We also denote the extremal cost process, generating the filtration $\mathcal{N}_\tau$, as $\hat{f}_\tau$. Within a deterministic setting, a smooth cost functional $f(\mathbf{x})$ tends to become stationary (in the sense of a vanishing first variation) as $\mathbf{x}$ approaches an extremal value $\mathbf{x}^*$. In the stochastic setup adopted here, a counterpart of this scenario would be that any conditioning of the process $\mathbf{x}_\tau$ during a future iteration on $\mathcal{N}_\xi$, which is the filtration generated by $\hat{f}_s$, $s < \xi$, till a past iteration (i.e. $\xi < \tau$), identically yields the random variable $\mathbf{x}_\xi$ itself, i.e. $E(\mathbf{x}_\tau | \mathcal{N}_\xi) = \mathbf{x}_\xi$. Using the Markov structure of $\mathbf{x}_\tau$, one may thus postulate that a necessary and sufficient condition for local extremization of $f_\tau := f(\mathbf{x}_\tau)$ is to require $E(\mathbf{x}_\tau | \hat{f}_\xi) = \mathbf{x}_\xi$. Interestingly, this characterization endows $\mathbf{x}_\tau$ with the martingale property with respect to the extremal cost filtration $\mathcal{N}_\tau$, i.e. once locally extremized, any future conditional mean of $\mathbf{x}_\tau$ on the cost filtration remains iteration invariant. We refer to [17] for an introductory treatise on the theory of martingales. An equivalent way of stating this is through a so called innovation process defined as $\hat{f}_\tau - f_\tau$, wherein local extremization would require driving the process $\mathbf{x}_\tau$ for an extremal process $\mathbf{x}_\tau^*$ such that $\hat{f}_\tau - f(\mathbf{x}_\tau^*)$ becomes a zero mean martingale, e.g. a zero-mean Brownian motion or, more generally, a stochastic integral in Ito's sense (or, in a $\tau$-discrete setting, a zero-mean random walk). Thus the determination of the local extrema is herein posed as a martingale problem, as originally conceptualized by Stroock and Varadhan [14] to provide a general setting for solutions to SDEs. If the optimization scheme is effective, one anticipates $\mathbf{x}_\tau \to \mathbf{x}_\tau^*$ with increasing iterations, i.e. as $\tau \to \infty$. Moreover, as the noise intensity associated with $\mathbf{x}_\tau$ approaches zero, $\mathbf{x}_\tau^* \to \mathbf{x}^*$. However, since a strictly zero noise intensity is infeasible within the stochastic setup, the solution in a deterministic setting may be thought of as a degenerate version (with a Dirac measure) of that obtained through the stochastic scheme.

Note that a ready extension of this approach is possible with multi-objective cost functions, which would require building up the innovation as a vector-valued process. Similarly, the approach may also be adapted for treating equality constraints, wherein zero-valued constraint functions may be used to construct the innovation processes. The easy adaptability of the current setup for multiple cost functionals or constraints may be viewed as an advantage over most existing evolutionary optimization schemes, where a single cost functional needs to be constructed based on all the constraints, a feature that may possibly engender instability for a class of large dimensional problems.

Driving $\mathbf{x}_\tau$ to an $\mathcal{N}_\tau$-martingale, or forcing the innovation process to a zero-mean martingale, may be accomplished expeditiously through a change of measures. In evolutionary algorithms, e.g. the GA, the accompanying change of measures is iteratively attained by assigning weights or fitness values to the current set of particles and subsequently selecting those with higher fitness, say, via rejection sampling. In an effort to explore the sample space better, these methods often use steps like 'crossover' and 'mutation'. These steps may be considered as biologically inspired safeguards against an intrinsic limitation of a weight-based approach, which tends to diminish all the weights but one to zero, thereby leaving the realized particles nearly identical (the problem of weight collapse or particle impoverishment) and thus precipitating premature convergence to a wrong solution. A more efficacious strategy to resist such degeneracy in the realized population set (i.e. the ensemble) is devised in the proposed scheme, wherein $\mathbf{x}_\tau$ at $\tau = \tau_i$, is updated using a purely additive term, derived through a Girsanov change of measures $P \to Q$ (see [14] for a detailed exposition on the Girsanov transformation), so as to ensure that the innovation process, originally described using measure $P$, becomes a zero-mean martingale under the new measure $Q$. A basic ingredient in effecting this change of measure is the Radon-Nikodym derivative $\Lambda_\tau := \frac{dP}{dQ}$ (assuming absolute continuity of $Q$ w.r.t. $P$ and vice versa), a scalar valued random variable also called the likelihood ratio that weighs a particle $\mathbf{x}_\tau$ as $\mathbf{x}_\tau \Lambda_\tau$. In executing the search for local extrema, the proposed scheme obtains the additive particle update by expanding

$\mathbf{x}_\tau \Lambda_\tau$ using Ito's formula [14]. An immediate advantage of using the additive update is that the particles with lower weights are never 'killed', but rather corrected to become more competitive by being driven closer to the local optima. In direct analogy with Taylor's expansion of a smooth function(al), where the first order term is based on Newton's directional derivative, the current version of the additive correction term may be thought of as a non-Newton directional term that drives the innovation process to a zero-mean martingale and a precise form of this term is derived in Section 3. As noted before, the innovation process may be a vector; such a scenario may even include attempts at addressing possible numerical instability for a class of optimization problems by splitting a single cost functional into several and then driving each of the correspondent innovation into a zero-mean martingale so as to ensure that the innovation corresponding to the original cost functional is also driven to a zero-mean martingale. Thus, for a simpler exposition without losing generality, the building blocks of the method are presented using a single cost functional only. We emphasize that the formulation is trivially adapted for vector innovation processes of any finite dimension.

Within a $\tau$-discretized framework and with $\tau \in (\tau_{i-1}, \tau_i]$, the evolution of $\mathbf{x}_\tau$ follows Eqn. 2.1. The innovation constraint that must be satisfied for the local search is given by:

$$\hat{f}_\tau - f(\mathbf{x}_\tau) = \Delta \eta_\tau \qquad (2.3a)$$

where $\hat{f}_\tau \in \mathbb{R}$ is the extremal cost process. $f$ the non-linear (possibly non-smooth) cost functional and $\Delta \eta_\tau = \eta_\tau - \eta_{i-1} \in \mathbb{R}$ a $P$-Brownian increment representing the diffusive fluctuations. Since the derivation of the subsequent integro-differential equation for the local search is conveniently accomplished by imparting to Eqn. 2.3(a) the explicit form of an SDE, an $\mathcal{N}_\tau$-measureable process $\breve{f}_\tau := \breve{f}(\tau)$ may be constructed to arrive at the following incremental form:

$$\Delta \breve{f}_\tau := \hat{f}_\tau \Delta \tau = f(\mathbf{x}_\tau)\Delta \tau + \Delta \eta_\tau \Delta \tau; \quad \Delta \tau = \tau - \tau_{i-1} \qquad (2.3b)$$

Since the $\tau$-axis is entirely fictitious, $\Delta \tau_i := \tau_i - \tau_{i-1}$ is taken to be a 'small' increment. Hence, replacing $\Delta \tau$ with $\Delta \tau_i = \tau_i - \tau_{i-1}$, Eqn. 2.3(b) may be recast as:

$$\Delta \breve{f}_\tau = f(\mathbf{x}_\tau)\Delta\tau + \Delta\eta_\tau \Delta\tau_i \qquad (2.3c)$$

which is essentially correspondent to the SDE:

$$d\breve{f}_\tau = f(\mathbf{x}_\tau)d\tau + \Delta\tau_i d\eta_\tau \qquad (2.3d)$$

Note that the replacement of $\Delta\tau$ by $\Delta\tau_i$ merely modifies the intensity of the noise process $\Delta\eta_\tau$ in Eqn. 2.3(a) and does not, in any way, interfere with the basic goal of driving the innovation constraint to a zero-mean martingale. Indeed the form of the SDE 2.3(d) implies that the diffusion coefficient $\Delta\tau_i$ is an order 'smaller' relative to the drift coefficient $f(\mathbf{x}_\tau)$. However, since $\eta_\tau$ is not a standard Brownian motion, it is more convenient to rewrite Eqn. 2.3(d) as:

$$d\breve{f}_\tau = f(\mathbf{x}_\tau)d\tau + \rho_\tau dW_\tau \qquad (2.3e)$$

where $W_\tau$ is a standard *P*-Brownian motion and $\rho_\tau$ is a more general form of (scalar-valued) noise intensity that may explicitly depend on $\tau$. However for multi-objective optimization problems that involve more than one cost functionals, $\rho_\tau$ could be an intensity matrix and hence, for the sake of a general treatment, would be considered as such in further manipulations. Eqn. 2.3(e) may finally be rewritten as:

$$d\tilde{f}_\tau := \rho_\tau^{-1} d\breve{f}_\tau = h(\mathbf{x}_\tau)d\tau + dW_\tau$$

where $h(\mathbf{x}_\tau) := \rho_\tau^{-1} f(\mathbf{x}_\tau)$. While it is possible, or even desirable, to replace the Brownian (or diffusion-type) noise term above by one whose quadratic variation is zero (e.g. a Poisson type noise), such a modification is not central to the basic idea and will be considered in future extensions of the work. The SDE 2.3(e) is assumed to satisfy the standard existence criteria [14] so that it is at least weakly solvable. The $\mathcal{N}_\tau$-measurable locally optimal solution may now be identified with the conditional mean $E(\mathbf{x}_\tau | \mathcal{N}_\tau)$, a measure-valued process characterized through the conditional distribution of $\mathbf{x}_\tau$ given the extremal cost process until $\tau$. Considering a new measure $Q$ under which $\mathbf{x}_\tau$ from Eqn. (2.1) satisfies the constraint 2.3(a), the conditional mean may be represented via the generalized Bayes' formula as:

$$\pi_\tau(\mathbf{x}) := E(\mathbf{x}_\tau | \mathcal{N}_\tau) = \frac{E_Q(\mathbf{x}_\tau \Lambda_\tau | \mathcal{N}_\tau)}{E_Q(\Lambda_\tau | \mathcal{N}_\tau)} \tag{2.4}$$

where the expectation $E_Q(\cdot)$ is taken with respect to the new measure $Q$ and $\Lambda_\tau$ is the likelihood given by:

$$\Lambda_\tau = \exp\left[\int_{\tau_{i-1}}^{\tau} h_s d\tilde{f}_s - \frac{1}{2}\int_{\tau_{i-1}}^{\tau} h_s^2 ds\right] \tag{2.5}$$

As derived in **Appendix I** via manipulations on $\Lambda_\tau$, the incrementally additive updates on $\mathbf{x}_\tau$ to arrive at the local extrema must be consistent with the following differential equation:

$$d\pi_\tau(\mathbf{x}) = \left(\pi_\tau(\mathbf{x}f) - \pi_\tau(\mathbf{x})\pi_\tau(f)\right)\left(\rho_\tau \rho_\tau^T\right)^{-1}\left(d\breve{f}_\tau - \pi_\tau(f)d\tau\right) \tag{2.6}$$

Here, $\left(d\breve{f}_\tau - \pi_\tau(f)d\tau\right)$ is the incremental innovation process which is driven to a zero-mean martingale. An equivalent integral representation of Eqn. 2.6, herein referred to as the extremal equation, is:

$$\pi_\tau(\mathbf{x}) = \pi_{i-1}(\mathbf{x}) + \int_{i-1}^{\tau}\left(\pi_s(\mathbf{x}f) - \pi_s(\mathbf{x})\pi_s(f)\right)\left(\rho_s \rho_s^T\right)^{-1}\left(d\breve{f}_s - \pi_s(f)ds\right) \tag{2.7}$$

In fact, the appearance of the unknown term $\pi_\tau(f)$ on the right hand side (RHS) prevents Eqn. 2.6 or 2.7 to be considered as an SDE in $\pi_\tau(\mathbf{x})$. Indeed, a necessarily nonlinear dependence of the cost functional $f$ on $\mathbf{x}_\tau$ and the consequent non-Gaussianity of $\pi_\tau(f)$ would prevent writing the latter in terms of $\pi_\tau(\mathbf{x})$, thereby leading to the so called closure problem in solving for $\pi_\tau(\mathbf{x})$.

## 3. Discretization of the extremal equation

While a direct solution of Eqn. 2.6 or 2.7 yields the local extrema in principle, exact/analytical solutions may be ruled out owing to the circularity inherent in the closure problem. This has a parallel in the theory of nonlinear stochastic filtering, wherein the Kushner-Stratonovich equation [18] (an equivalent of Eqn. 2.6) also suffers from a similar circularity problem.

Motivated by the MC filters often used to solve nonlinear filtering problems [19, 20], an MC scheme may similarly be developed for a numerical treatment of Eqn. 2.6 or 2.7 as well. Specifically, a two stage strategy, viz. prediction and update, may apparently be considered, even though, as we will soon see, the prediction step could be entirely eliminated in a variant of the final scheme. As with most evolutionary optimization schemes, a random exploration (prediction) may first be undertaken over the $i^{\text{th}}$ iteration, i.e. over $(\tau_{i-1}, \tau_i]$, based on Eqn. 2.1. Denoting the ensemble size by $N$, one thus realizes $N$ predicted particles or MC candidates, $\{\mathbf{x}_\tau^{(j)}\}_{j=1}^N$ that must be updated based on Eqn. 2.7. In developing an MC-based numerical solution to Eqn. 2.7, a sample-averaged version of the equation is first written as:

$$\pi_\tau^N(\mathbf{x}) = \pi_{i-1}^N(\mathbf{x}) + \int_{\tau_{i-1}}^{\tau} \left( \pi_s^N(\mathbf{x}f) - \pi_s^N(\mathbf{x})\pi_s^N(f) \right) \left( \rho_s \rho_s^T \right)^{-1} \left( d\breve{f}_s - \pi_s^N(f)ds \right) \tag{3.1}$$

$\pi^N(.) = (1/N) \sum_{j=1}^N (.)^{(j)}$ is the ensemble-averaged approximation to the conditional mean $\pi(.)$. A candidate-wise representation of Eqn. 3.1 may be given by:

$$\mathbf{X}_\tau = \mathbf{X}_{i-1} + \frac{1}{N} \int_{\tau_{i-1}}^{\tau} \left\{ \mathbf{X}_s \mathbf{F}_s^T - \hat{\mathbf{X}}_s \hat{\mathbf{F}}_s^T \right\} \left( \rho_s \rho_s^T \right)^{-1} \left\{ d\breve{F}_s - \mathbf{F}_s ds \right\} \tag{3.2}$$

where $\mathbf{X}_\tau := [\mathbf{x}_\tau^{(1)}, \mathbf{x}_\tau^{(2)}, ... \mathbf{x}_\tau^{(N)}]$, $\mathbf{F}_\tau := [f_\tau^{(1)}, f_\tau^{(2)}, ..., f_\tau^{(N)}]$, $\hat{\mathbf{X}}_\tau := \pi_\tau^N(\mathbf{x})\mathbf{r} \in \mathbb{R}^{n \times N}$, $\hat{\mathbf{F}}_\tau := \pi_\tau^N(f)\mathbf{r} \in \mathbb{R}^N$, $d\breve{F}_\tau := d\breve{f}_s \mathbf{r} \in \mathbb{R}^N$. $\mathbf{r} = \{1, 1, ..., 1\} \in \mathbb{R}^N$ is an $N$-dimensional row vector with entries 1.

Note that the second term on the RHS of Eqn. 3.2 is the update/correction term that has its parallel with the directional derivative term in gradient based updates involving a smooth functional. For solving Eqn. 3.2, the MC approximation for Eqn. 2.7, a $\tau$-discrete numerical scheme is required. Such a scheme would typically involve the following two steps.

*a) Prediction*

The predicted candidate set $\tilde{\mathbf{X}}_\tau = [\tilde{\mathbf{x}}_\tau^{(1)}, ..., \tilde{\mathbf{x}}_\tau^{(N)}]$ at $\tau$ are generated using the following Euler Maruyama (EM) disctretized map:

$$\tilde{\mathbf{X}}_\tau = \mathbf{X}_{i-1} + \Delta \boldsymbol{\psi}_\tau \tag{3.3}$$

where $\Delta \boldsymbol{\psi}_\tau := [\Delta \xi_\tau^{(1)},...,\Delta \xi_\tau^{(N)}]$, $\Delta \xi_\tau^{(1)} = \xi_\tau^{(1)} - \xi_{i-1}^{(1)}$ etc.

*b) Additive Update*

The predicted candidates are improved using the correction term as in Eqn. 3.2 based on an EM approximation to the integral. In both prediction and update, explicit EM schemes have been used merely for expositional convenience, even though other integration schemes could be applicable, especially for integrating the correction integral. The discrete update equation is given by:

$$\mathbf{X}_\tau = \tilde{\mathbf{X}}_\tau + \frac{1}{N}\left\{\tilde{\mathbf{X}}_\tau \tilde{\mathbf{F}}_\tau^T - \hat{\tilde{\mathbf{X}}}_\tau \hat{\tilde{\mathbf{F}}}_\tau^T\right\}\left(\rho_\tau \rho_\tau^T\right)^{-1}\left\{\Delta \breve{F}_\tau - \tilde{\mathbf{F}}_\tau \Delta \tau\right\} \tag{3.4}$$

Here $\Delta \breve{F}_\tau = \Delta \breve{f}_\tau \mathbf{r}$, $\tilde{\mathbf{F}}_\tau := [\tilde{f}_\tau^{(1)},...,\tilde{f}_\tau^{(N)}]$ and the predicted candidates, $\tilde{\mathbf{x}}_\tau$, are used to compute $\tilde{f}_\tau := f(\tilde{\mathbf{x}}_\tau)$. Also note that $\hat{\tilde{\mathbf{F}}}_\tau := \pi_\tau^N(\tilde{f})\mathbf{r}$ and $\hat{\tilde{\mathbf{X}}}_\tau := \pi_\tau^N(\tilde{\mathbf{x}})\mathbf{r}$. It is convenient to recast Eqn. 3.4 as:

$$\mathbf{X}_\tau = \tilde{\mathbf{X}}_\tau + \frac{1}{N}\left\{\left(\tilde{\mathbf{X}}_\tau - \hat{\tilde{\mathbf{X}}}_\tau\right)\tilde{\mathbf{F}}_\tau^T + \hat{\tilde{\mathbf{X}}}_\tau\left(\tilde{\mathbf{F}}_\tau - \hat{\tilde{\mathbf{F}}}_\tau\right)^T\right\}\left(\rho_\tau \rho_\tau^T\right)^{-1}\left\{\Delta \breve{F}_\tau - \tilde{\mathbf{F}}_\tau \Delta \tau\right\} \tag{3.5}$$

From Eqn. 2.3(b), recall that $\Delta \breve{f}_\tau := \hat{f}_\tau \Delta \tau$ and thus Eqn. 3.5 may be rearranged as:

$$\mathbf{X}_\tau = \tilde{\mathbf{X}}_\tau + \frac{1}{N}\left\{\left(\tilde{\mathbf{X}}_\tau - \hat{\tilde{\mathbf{X}}}_\tau\right)\left(\tilde{\mathbf{F}}_\tau^T \Delta \tau\right) + \left(\hat{\tilde{\mathbf{X}}}_\tau \Delta \tau\right)\left(\tilde{\mathbf{F}}_\tau - \hat{\tilde{\mathbf{F}}}_\tau\right)^T\right\}\left(\rho_\tau \rho_\tau^T\right)^{-1}\left\{\hat{\mathbf{F}}_\tau - \tilde{\mathbf{F}}_\tau\right\} \tag{3.6}$$

where $\hat{\mathbf{F}}_\tau = \hat{f}_\tau \mathbf{r} \in \mathbb{R}^N$. When the candidate solutions are far away from the local extrema, the correction terms should be large such that the candidates traverse more in the search space. In such cases, as during the initial stages of evolution, the innovation process may be far from behaving like a zero-mean martingale, i.e. it may have a significant drift component and the gain-like coefficient matrix in the correction term should enable better exploration of the sample space. (e.g. by having a large norm). Since the estimates in this regime may have sharper gradients in $\tau$, one way to modify the coefficient matrix in Eqn. 3.6 would be to incorporate information on these gradients through the previous estimates. Towards this, $\tilde{\mathbf{F}}_\tau^T \Delta \tau$ and $\tilde{\mathbf{X}}_\tau \Delta \tau$ are replaced respectively by the following approximations:

$$\tilde{\mathbf{F}}_\tau^T \Delta \tau \approx \left( \tilde{\mathbf{F}}_\tau^T \tau - \tilde{\hat{\mathbf{F}}}_{i-1}^T \tau_{i-1} - \Delta \tilde{\hat{\mathbf{F}}}_\tau^T \tau \right) \tag{3.7a}$$

$$\tilde{\hat{\mathbf{X}}}_\tau \Delta \tau \approx \left( \tilde{\hat{\mathbf{X}}}_\tau \tau - \tilde{\hat{\mathbf{X}}}_{i-1} \tau_{i-1} \right) \tag{3.7b}$$

Note that we have used Ito's formula while approximating $\tilde{\mathbf{F}}_\tau^T \Delta \tau$ in Eqn. 3.7(a). Using Eqn. 3.7(a-b), Eqn. 3.6 may be modified as:

$$\mathbf{X}_\tau = \tilde{\mathbf{X}}_\tau + \frac{1}{N} \left\{ \begin{array}{l} \left( \tilde{\mathbf{X}}_\tau - \tilde{\hat{\mathbf{X}}}_\tau \right) \left( \tilde{\mathbf{F}}_\tau^T \tau - \tilde{\hat{\mathbf{F}}}_{i-1}^T \tau_{i-1} - \Delta \tilde{\hat{\mathbf{F}}}_\tau^T \tau \right) \\ + \left( \tilde{\hat{\mathbf{X}}}_\tau \tau - \tilde{\hat{\mathbf{X}}}_{i-1} \tau_{i-1} \right) \left( \tilde{\mathbf{F}}_\tau - \tilde{\hat{\mathbf{F}}}_\tau \right)^T \end{array} \right\} \left( \rho_\tau \rho_\tau^T \right)^{-1} \left\{ \widehat{\mathbf{F}}_\tau - \tilde{\mathbf{F}}_\tau \right\} \tag{3.8}$$

It may also be observed that, once the conditional mean is obtained, the innovation noise covariance $\rho_\tau \rho_\tau^T$ should satisfy the following identity:

$$\rho_\tau \rho_\tau^T \approx \pi_\tau^N \left( \left( \widehat{f}_\tau - \tilde{f}_\tau \right) \left( \widehat{f}_\tau - \tilde{f}_\tau \right)^T \right) = \frac{1}{N-1} \left( \left( \widehat{\mathbf{F}}_\tau - \tilde{\mathbf{F}}_\tau \right) - \left( \widehat{\mathbf{F}}_\tau - \tilde{\hat{\mathbf{F}}}_\tau \right) \right) \left( \left( \widehat{\mathbf{F}}_\tau - \tilde{\mathbf{F}}_\tau \right) - \left( \widehat{\mathbf{F}}_\tau - \tilde{\hat{\mathbf{F}}}_\tau \right) \right)^T$$

One anticipates that, away from the local extrema, the (norm of the) RHS of the equation above should typically be relatively large. Accordingly, in order to impart higher diffusion to the local search in the initial stages wherein the innovation could have a significant drift component, $\rho_\tau^T \rho_\tau$ may be replaced by:

$$\alpha \frac{1}{N-1} \left( \left( \widehat{\mathbf{F}}_\tau - \tilde{\mathbf{F}}_\tau \right) - \left( \widehat{\mathbf{F}}_\tau - \tilde{\hat{\mathbf{F}}}_\tau \right) \right) \left( \left( \widehat{\mathbf{F}}_\tau - \tilde{\mathbf{F}}_\tau \right) - \left( \widehat{\mathbf{F}}_\tau - \tilde{\hat{\mathbf{F}}}_\tau \right) \right)^T + (1-\alpha) \rho_\tau \rho_\tau^T$$

in Eqn. 3.8. Here $0 < \alpha < 1$ (typically taken as 0.8 in the numerical illustrations provided later on). Finally, then, Eqn. 3.8 may be written as:

$$\mathbf{X}_\tau = \tilde{\mathbf{X}}_\tau + \frac{1}{N}\left\{\left(\tilde{\mathbf{X}}_\tau - \tilde{\tilde{\mathbf{X}}}_\tau\right)\left(\tilde{\mathbf{F}}_\tau^T \tau - \tilde{\mathbf{F}}_{i-1}^T \tau_{i-1} - \Delta\tilde{\mathbf{F}}_\tau^T \tau\right) + \left(\tilde{\tilde{\mathbf{X}}}_\tau \tau - \tilde{\tilde{\mathbf{X}}}_{i-1}\tau_{i-1}\right)\left(\tilde{\mathbf{F}}_\tau - \tilde{\tilde{\mathbf{F}}}_\tau\right)^T\right\}$$

$$\left(\alpha \frac{1}{N-1}\left(\left(\hat{\mathbf{F}}_\tau - \tilde{\mathbf{F}}_\tau\right) - \left(\hat{\mathbf{F}}_\tau - \tilde{\tilde{\mathbf{F}}}_\tau\right)\right)\left(\left(\hat{\mathbf{F}}_\tau - \tilde{\mathbf{F}}_\tau\right) - \left(\hat{\mathbf{F}}_\tau - \tilde{\tilde{\mathbf{F}}}_\tau\right)\right)^T + (1-\alpha)\rho_\tau \rho_\tau^T\right)^{-1}\left\{\hat{\mathbf{F}}_\tau - \tilde{\mathbf{F}}_\tau\right\}$$

(3.9)

In a more concise form, the update equation is thus of the form:

$$\mathbf{X}_\tau = \tilde{\mathbf{X}}_\tau + \tilde{\mathbf{G}}_\tau\left\{\hat{\mathbf{F}}_\tau - \tilde{\mathbf{F}}_\tau\right\} \tag{3.10}$$

where the gain-like update coefficient matrix is given by:

$$\tilde{\mathbf{G}}_\tau := \frac{1}{N}\left\{\left(\tilde{\mathbf{X}}_\tau - \tilde{\tilde{\mathbf{X}}}_\tau\right)\left(\tilde{\mathbf{F}}_\tau^T \tau - \tilde{\mathbf{F}}_{i-1}^T \tau_{i-1} - \Delta\tilde{\mathbf{F}}_\tau^T \tau\right) + \left(\tilde{\tilde{\mathbf{X}}}_\tau \tau - \tilde{\tilde{\mathbf{X}}}_{i-1}\tau_{i-1}\right)\left(\tilde{\mathbf{F}}_\tau - \tilde{\tilde{\mathbf{F}}}_\tau\right)^T\right\}$$

$$\left(\alpha \frac{1}{N-1}\left(\left(\hat{\mathbf{F}}_\tau - \tilde{\mathbf{F}}_\tau\right) - \left(\hat{\mathbf{F}}_\tau - \tilde{\tilde{\mathbf{F}}}_\tau\right)\right)\left(\left(\hat{\mathbf{F}}_\tau - \tilde{\mathbf{F}}_\tau\right) - \left(\hat{\mathbf{F}}_\tau - \tilde{\tilde{\mathbf{F}}}_\tau\right)\right)^T + (1-\alpha)\rho_\tau \rho_\tau^T\right)^{-1}$$

In the update strategy described in Eqn. 3.10, while some measures are already taken for an effective exploration of the search space, we may still improve the exploration by adopting an annealing type approach. The idea is to provide larger diffusion intensity to the update term in the initial stages of evolution and reduce it as the candidates approach the global optima (a proper search scheme for the global optima is presented in the next section). The annealing-type coefficient $\beta_\tau$ (with $1/\beta_\tau$ interpreted as the annealing temperature) here appears as a scalar factor multiplying the update term so that the update equation becomes:

$$\mathbf{X}_\tau = \tilde{\mathbf{X}}_\tau + \beta_\tau \tilde{\mathbf{G}}_\tau\left\{\hat{\mathbf{F}}_\tau - \tilde{\mathbf{F}}_\tau\right\} \tag{3.11}$$

Typically, $\beta_\tau$ starts with a small positive (near-zero) value at $\tau = \tau_0$ (implying that the initial temperature $1/\beta_0$ is high) and gradually increases to $\beta_{max}$ at $\tau_{max}$, the end of the iterations. In the numerical simulations presented here, $\beta_{max}$ has been taken as 2. $\beta_\tau$ is analogous to the annealing parameter found in the annealed Markov chain Monte Carlo (MCMC) simulation. Although the annealing temperature in the MCMC framework is prescribed to be reduced very slowly (e.g. logarithmically), $1/\beta_\tau$ in the present scheme may be reduced much faster (e.g. exponentially) in view of the fact that the job of an efficient exploration of the search space is already attended to, in part, through an appropriate construction of the coefficient matrix $\tilde{\mathbf{G}}_\tau$. Indeed, while virtually any form of $\beta_\tau$ as an increasing function of $\tau$ may be prescribed, we presently use $\beta_{\tau_i} = \beta_i = \beta_{\max} \exp(i+1-\kappa)$ within a $\tau$-discrete setting, where $\kappa$ denotes the total number of iterations, i.e. $\beta_\kappa := \beta_{\max} = \beta_{\tau_{\max}}$.

## 4. Coalescence and scrambling: schemes for global search

In any global optimization scheme, a major challenge is to eliminate the possibility of the candidate solutions getting trapped in local extrema. Although it seems possible to avoid the local traps through an exploitation of the innovation noise, whose intensity could be tuned using the annealing-type factor $\beta_\tau$, such an approach could be quite inefficient. Indeed, as a local extremal point is approached, the 'strength' (or norm) of the update term (which is analogous to a directional derivative term in a gradient-based search) would be small and consequently the sensitivity of the update term to variations in $\beta_\tau$ would also be poor. This makes choosing $\beta_\tau$ difficult (e.g. necessitating $\beta_\tau$ to be too small for the search scheme to be efficient) and renders the annealing-type scheme less effective for the global search. Moreover, smaller $\beta_\tau$ also implies larger diffusion and hence poorer convergence. A more effective way is probably to randomly perturb the candidates, so that they do not get trapped in the local wells. One such perturbation scheme, combining two approaches that are herein referred to as 'coalescing' and 'scrambling', may be accomplished respectively via yet another martingale problem and a perturbation kernel as explained below. Before going into the details of 'coalescing', we recall that the local extremization, posed as a martingale problem, requires that the $j^{\text{th}}$ candidate should be updated following Eqn. 3.11 as:

$$\mathbf{x}_i^{(j)} = \tilde{\mathbf{x}}_i^{(j)} + \tilde{\mathbf{C}}_i^{(j)} \tag{4.1}$$

where $\tilde{\mathbf{C}}_i^{(j)} := \beta_i \tilde{\mathbf{G}}_i \left\{ \hat{f}_i - \tilde{f}_i^{(j)} \right\}$. Being essentially a scheme for local extremization with the annealing-type term playing a poor role in the global search, the global update strategy must be efficiently equipped to prevent the evolving candidates from getting trapped at the local wells, i.e. the 'bad points'. In doing so, the basic idea here is to provide several layers of random perturbation to the candidate solutions, whose intensity may be formally thought of as being indexed by a positive integer $l$ such that the perturbation vanishes as $l \to \infty$. Within the $\tau$-discrete setting, we start with the prediction $\tilde{\mathbf{x}}_i$ and denote $^l\Delta\hat{\mathbf{x}}_i = {}^l\mathbf{x}_i - {}^l\tilde{\mathbf{x}}_i$ as the randomly perturbed, $l$-indexed increment so that its limiting dynamics is provided by the process $^\infty\Delta\hat{\mathbf{x}}_i \to \Delta\mathbf{x}_i$ as the random perturbations vanish asymptotically. During the $i^{th}$ iteration, the perturbed increment is arrived using two transitional increments, $^l\mathbf{u}_i$ and $^l\mathbf{v}_i$ representing populations corresponding to the two perturbation operators, say **T1** and **T2** respectively. While the operator **T1** corresponds to the combined 'local search' and 'coalescence' operations, **T2** denotes the 'scrambling' operation. Details of both 'coalescence' and 'scrambling' operations will be provided shortly. Specifically, the transitions may be written as, $^l\Delta\mathbf{x}_i \xrightarrow{\mathbf{T1}} {}^l\mathbf{u}_i \xrightarrow{\mathbf{T2}} {}^l\mathbf{v}_i \xrightarrow{\mathbf{T3}} {}^l\Delta\hat{\mathbf{x}}_i$ where **T3** is a selection operator, commonly used with most evolutionary optimization schemes in some form or the other. Here $^l\Delta\hat{\mathbf{x}}_i$ is the finally obtained solution increment at $\tau_i$ which goes as input to the next iteration, i.e. $^l\mathbf{x}_i := {}^l\tilde{\mathbf{x}}_i + {}^l\Delta\hat{\mathbf{x}}_i$ and $^l\Delta\mathbf{x}_i = {}^l\tilde{\mathbf{x}}_i - {}^l\mathbf{x}_{i-1}$ is the predicted increment. Ideally, one may start the iterations with a small $l$ (i.e. high perturbation intensity) and gradually increase $l$ with progressing iterations. However, in the presented numerical implementation of our scheme, we keep the perturbation intensity uniformly small all through the iterations. We can manage to take this latitude as our method arguably has superior convergence features vis-a-vis most existing evolutionary optimization algorithms, as demonstrated, to a limited extent, in Section 5. In view of this and for notational ease, the left superscript $l$ is often removed from the variables in the discussion to follow (provided that the perturbed nature of the variables is clear from the context). The operators are now defined below.

**Operator T1**: **Local search and coalescence**

The operation for the local search has already been described and quantitatively captured through Eqn. 4.1. Thus, in the following, we discuss the operation of 'coalescence'. This perturbation is motivated by the observation that the probability density function (PDF, if it exists) associated with the converged measure $\pi_\tau(.)$ should be unimodal, with its only peak located at the global extremum. Thus, when convergence to the global extremum occurs, all the candidate solutions should coalesce at the globally optimum point, except for a zero-mean noisy scatter around the latter. Ideally, for the sake of optimization accuracy, the noisy scatter should also have a low intensity. Once the global optimization scheme converges, the noisy scatter should then behave as a zero-mean martingale as a function of $\tau$ and with a unimodal transitional PDF. A zero-mean Brownian motion, which has a Gussian PDF, is one such martingale. Clearly, such a property does not hold away from the global optimum, where the PDF should be multi-modal with a peak at every local extremum detected by the algorithm.

Now, we wish to make the above argument a little more precise and thus obtain a scheme to force the coalescence of candidates. Consider the update of the $j^{\text{th}}$ candidate $\mathbf{x}_\tau^{(j)}$ such that coalescence of candidates can be enforced. A measure of the random scatter around $\mathbf{x}_\tau^{(j)}$ could be defined as $\delta_\tau(j) = \mathbf{x}_\tau^{(\sigma_1(j))} - \mathbf{x}_\tau^{(j)}$, where $\sigma_1(j)$ denotes a random permutation on the indexing set $\{1, N\} \setminus \{j\}$ based on a uniform measure. Our goal is to drive $\delta_\tau(j) = \mathbf{x}_\tau^{(\sigma_1(j))} - \mathbf{x}_\tau^{(j)}$ to a zero-mean vector Brownian increment $\Delta \eta_\tau^\mathbf{c}$ with intensity matrix $\rho_\tau^\mathbf{c}$ (typically diagonal), which may be chosen uniformly for all $j$. The word 'coalescence' then implies that, in the limit of the noise term in $\delta_\tau(j)$ approaching zero, all the candidates would tend to coalesce into a single particle at the global extremum. Thus, similar to the innovation $\hat{f}_\tau - f(\mathbf{x}_\tau)$ on the left hand side (LHS) of Eqn. 2.3(a), one treats $\mathbf{x}_\tau^{(\sigma_1(j))} - \mathbf{x}_\tau^{(j)}$ as yet another innovation process, whose characterization upon convergence would be of the form $\mathbf{x}_\tau^{(\sigma_1(j))} - \mathbf{x}_\tau^{(j)} = \Delta \eta_\tau^\mathbf{c}$. Since $\Delta \eta_\tau^\mathbf{c}$ provides for an additional layer of randomness that imparts to each candidate $\mathbf{x}_\tau^{(j)}$ the structure of a stochastic process, the extremal cost filtration $\mathcal{N}_\tau$ may now be suitably expanded to include the sub-filtration generated by $\Delta \eta_s^\mathbf{c}$ for $s \leq \tau$ in order to remain theoretically consistent. Including the coalescence innovation within our search process, Eqn. 4.1 may be modified as:

$$\mathbf{x}_i^{(j)} = \tilde{\mathbf{x}}_i^{(j)} + \tilde{\mathbf{D}}_i^{(j)} \quad \text{or} \quad \mathbf{u}_i^{(j)} = \tilde{\mathbf{D}}_i^{(j)} \qquad (4.2a)$$

where $\tilde{\mathbf{D}}_i^{(j)} := \beta_i \tilde{\mathbf{G}}_i \tilde{\mathbf{I}}_i^{(j)}$, $\tilde{\mathbf{I}}_i^{(j)} := \left\{ \begin{array}{c} \hat{f}_i - \tilde{f}_i^{(j)} \\ \tilde{\mathbf{x}}_i^{(\sigma_1(j))} - \tilde{\mathbf{x}}_i^{(j)} \end{array} \right\}$ (4.2b)

and we recall that over-tildes indicate either the predicted candidates or functions evaluated using the predicted candidates, as appropriate. Allowing for a convenient notational abuse, we have retained the same notation for the gain-like update coefficient matrix $\tilde{\mathbf{G}}_i$ in Eqn. 4.2(b) (used earlier in the local update Eqn. 3.10), which may now be computed as:

$$\tilde{\mathbf{G}}_i := \frac{1}{N} \left\{ \left( \tilde{\mathbf{X}}_i - \bar{\tilde{\mathbf{X}}}_i \right) \left( \tilde{\mathbf{F}}_i^T \tau_i - \tilde{\mathbf{F}}_{i-1}^T \tau_{i-1} - \Delta \bar{\tilde{\mathbf{F}}}_i^T \tau_i \right) + \left( \bar{\tilde{\mathbf{X}}}_i \tau_i - \bar{\tilde{\mathbf{X}}}_{i-1} \tau_{i-1} \right) \left( \tilde{\mathbf{F}}_i - \bar{\tilde{\mathbf{F}}}_i \right)^T \right\}$$

$$\left( \alpha \frac{1}{N-1} \left( \left( \hat{\mathbf{F}}_i - \tilde{\mathbf{F}}_i \right) - \left( \hat{\mathbf{F}}_i - \bar{\tilde{\mathbf{F}}}_i \right) \right) \left( \left( \hat{\mathbf{F}}_i - \tilde{\mathbf{F}}_i \right) - \left( \hat{\mathbf{F}}_i - \bar{\tilde{\mathbf{F}}}_i \right) \right)^T + (1-\alpha) \gamma_i \gamma_i^T \right)^{-1}$$

$$\gamma_i \gamma_i^T := \begin{bmatrix} \rho_i \rho_i^T & \mathbf{0} \\ \mathbf{0} & \rho_i^\mathbf{c} (\rho_i^\mathbf{c})^T \end{bmatrix}$$

Here the $j^{\text{th}}$ column of $\tilde{\mathbf{F}}_i$ is given as $\left\{ \begin{array}{c} \hat{f}_i - \tilde{f}_i^{(j)} \\ \tilde{\mathbf{x}}_i^{\sigma_1(j)} - \tilde{\mathbf{x}}_i^{(j)} \end{array} \right\}$. Let $a$ is a positive real number, define the closest integer smaller than $a$ by $\lfloor a \rfloor$. Then the integer valued perturbation parameter for the local extremization cum coalescence step may be identified as $l = \lfloor \| [\gamma_i \gamma_i^T]^{-1} \| \rfloor$. While we do not provide such a proof here, the convergence and uniqueness of the iterative increment through this step, for a non-decreasing sequence of $l$ converging to a limit point $l^*$ which may be large yet finite or infinity, may be shown based on the work of Stroock and Varadhan [15].

**Operator T2**: **Scrambling**

Similar to **T1**, the second perturbation operator is also based on random perturbations of the candidates in the population $\Omega$. Recalling that the gain-like coefficient matrix in the local update equation 4.1 is a derivative-free stochastic counterpart to the Frechet derivative, the term $\tilde{\mathbf{C}}_i^{(j)}$ may be considered the equivalent of the directional derivative term responsible for updating the $j^{th}$ candidate. Consequently, around any local extremum, the $L^2(P)$ norm $\|\tilde{\mathbf{C}}_i^{(j)}\|$ is likely to be small. This may render further updates of the $j^{th}$ particle trivially small, leading to a possible stalling of the local scheme. In order to move out of these local traps, the basic idea is to swap the gain-weighted directional information for the $j^{th}$ particle with that of another randomly selected one. This random perturbation at $\tau_i$ is accomplished by replacing the update equation 4.2(a) by:

$$\mathbf{x}_i^{(j)} = \tilde{\mathbf{x}}_i^{(j)} + \tilde{\mathbf{D}}_i^{\boldsymbol{\sigma}_2(j)}, \quad \text{or} \quad \mathbf{v}_i^{(j)} = \tilde{\mathbf{D}}_i^{\boldsymbol{\sigma}_2(j)} - \tilde{\mathbf{D}}_i^{(j)} \tag{4.6}$$

where $\tilde{\mathbf{D}}_i^{\boldsymbol{\sigma}_2(j)}$ is correction vector originally computed for the $\boldsymbol{\sigma}_2(j)^{th}$ candidate via Eqn. 4.2(b) and $\boldsymbol{\sigma}_2$ a random permutation on the integer set $\{1, N\}$. Such perturbation may be described by a probability kernel $p_l$ on $[1, N] \times [1, N]$ such that:

$$\sum_{i \in [1,N]} p_l(i, j) = 1 \ \forall j \in [1, N]$$

Clearly, as $l \to \infty$, the matrix $p_l(i, j)$ should ideally approach the identity matrix, i.e. $p_l(i, j) \to \delta_{ij}$, where $\delta_{ij}$ is the Kronecker delta. However, since the coalescence step ensures that all particles finally crowd around the mean of a unimodal PDF with progressing iterations, directional scrambling across the set of such converged particles should not, in any way, affect the numerical accuracy of the estimated global extremum. In other words, in practical implementations of our scheme, $p_l(i, j)$ need not strictly approach the identity matrix for large $l$.

**Operator T3: Selection**

Use of diffusion-based random perturbations during exploration may sometimes result in 'bad' candidates. This necessitates a selection step wherein candidates for the next iteration (say, the $i^{th}$ iteration) are chosen based on some selection criteria effected by a selection or fitness function

$g(\upsilon | \mathbf{x}_{i-1})$, $\upsilon \in \Omega$. A general construction of the function, which corresponds to a Markov transition kernel on $\Omega$ and is conditioned on the ensemble of particles for the last iteration, should satisfy the following properties:

a) $g(\upsilon = \mathbf{x}_i^{(j)} | \mathbf{x}_{i-1} = \mathbf{x}_{i-1}^{(k)}) = 0$ if $j \neq k$; $\forall j, k \in [1, N]$

b) $g(\upsilon = \mathbf{x}_i^{(j)} | \mathbf{x}_{i-1} = \mathbf{x}_{i-1}^{(j)}, f(\mathbf{x}_i^{(j)}) \geq f(\mathbf{x}_{i-1}^{(j)})) = \varsigma$ $\forall j \in [1, N]$ where $\varsigma \in (0, 1]$

c) $g(\upsilon = \mathbf{x}_{i-1}^{(j)} | \mathbf{x}_{i-1} = \mathbf{x}_{i-1}^{(j)}, f(\mathbf{x}_i^{(j)}) < f(\mathbf{x}_{i-1}^{(j)})) = \varsigma$

The updated $j^{\text{th}}$ particle $\mathbf{x}_i^{(j)}$ in the above clauses is computed using Eqn. 4.6, which combines all the three operations of local extremization, coalescence and scrambling. Here the integer-valued perturbation parameter $l$ may be identified with $l = \left\lfloor \dfrac{1}{1-\varsigma} \right\rfloor$. In the current numerical implementation of the scheme, we consistently take $\varsigma = 1$, which corresponds to $l$ being infinity across all iterations and implies that the selection procedure does not involve any random perturbation.

This work is primarily aimed at the proposal for a new evolutionary optimization scheme and a verification of its performance through the numerical route. Accordingly, a detailed convergence analysis of the asymptotic dynamics (i.e. as $l \to \infty$), based on a combination of the martingale theory of Stroock-Varadhan and the random perturbation theory of Freidlin-Wentzell, is left out of the scope of this work, even through a framework for accomplishing this task is laid out. For better clarity, the following pseudo-codes for the global optimization scheme should be helpful.

*The pseudo-code 1*:

1. Discretize the $\tau$-axis, say $[\tau_{\min}, \tau_{\max}]$, using a partition $\{\tau_0 = \tau_{\min}, \tau_1, ..., \tau_M = \tau_{\max}\}$ such that $\tau_0 < ... < \tau_M$ and $\tau_i - \tau_{i-1} = \Delta \tau_i$ $(= \dfrac{1}{M}$ if a uniform step size is chosen for $i = 0, ..., M-1)$. We assign $\tau_0 = 1$ and adopt $\Delta \tau$ small $(\sim 10^{-7})$. Choose an ensemble size $N$.

2. Generate the ensemble of initial population $\{\mathbf{x}_0^{(j)}\}_{j=1}^N$ for the solution vector. For each discrete $\tau_i$, $i=1,...,M-1$, execute the following steps.

3. *Prediction*

   Using $\{\mathbf{x}_{i-1}^{(j)}\}_{j=1}^N$, the last update (or the initial population for $i=1$) available at $\tau_{i-1}$, obtain the predicted candidates using:

   $$\tilde{\mathbf{x}}_i^{(j)} = \mathbf{x}_{i-1}^{(j)} + \Delta \boldsymbol{\xi}_i^{(j)}, \quad j=1,...,N$$

4. *Additive update*

   Choose $\alpha \in (0,1)$; a typically prescribed value would be $\alpha \approx 0.8$, even though the method also performs well for other values in the interval indicated.

   Update each particle using Eqn. 4.6, which is reproduced below:

   $$\mathbf{x}_i^{(j)} = \tilde{\mathbf{x}}_i^{(j)} + \tilde{\mathbf{D}}_i^{\boldsymbol{\sigma}_2(j)}, \quad j=1,...,N$$

   where $\tilde{\mathbf{D}}_i^{\boldsymbol{\sigma}_2(j)}$ is the $\boldsymbol{\sigma}_2(j)^{\text{th}}$ correction vector and $\boldsymbol{\sigma}_2(j)$ is the $j^{\text{th}}$ candidate from a random permutation based on a uniformly distributed measure on the integer set $\{1,...,N\}$. The expression for $\tilde{\mathbf{D}}_i^{\boldsymbol{\sigma}_2(j)}$ is also reproduced below:

   $$\tilde{\mathbf{D}}_i^{(j)} := \beta_i \tilde{\mathbf{G}}_i \tilde{\mathbf{I}}_i^{(j)}, \quad j=1,...,N,$$

   where $\beta_i = 1 - \dfrac{1}{\exp(i-1)}$

   and $\tilde{\mathbf{I}}_i^{(j)} := \left\{ \begin{array}{c} \hat{f}_i - \tilde{f}_i^{(j)} \\ \tilde{\mathbf{x}}_i^{\boldsymbol{\sigma}_1(j)} - \tilde{\mathbf{x}}_i^{(j)} \end{array} \right\}.$

   Recall that $\boldsymbol{\sigma}_1(j)$ is defined as the $j^{\text{th}}$ candidate from another (independently) random permutation of the integer set $\{1,...,N\}$.

$$\tilde{\mathbf{G}}_i := \frac{1}{N}\left\{\left(\tilde{\mathbf{X}}_i - \tilde{\hat{\mathbf{X}}}_i\right)\left(\tilde{\mathbf{F}}_i^T \tau_i - \tilde{\hat{\mathbf{F}}}_{i-1}^T \tau_{i-1} - \Delta\tilde{\hat{\mathbf{F}}}_i^T \tau_i\right) + \left(\tilde{\hat{\mathbf{X}}}_i \tau_i - \tilde{\hat{\mathbf{X}}}_{i-1}\tau_{i-1}\right)\left(\tilde{\mathbf{F}}_i - \tilde{\hat{\mathbf{F}}}_i\right)^T\right\}$$

$$\left(\alpha \frac{1}{N-1}\left(\left(\hat{\mathbf{F}}_i - \tilde{\mathbf{F}}_i\right) - \left(\hat{\mathbf{F}}_i - \tilde{\hat{\mathbf{F}}}_i\right)\right)\left(\left(\hat{\mathbf{F}}_i - \tilde{\mathbf{F}}_i\right) - \left(\hat{\mathbf{F}}_i - \tilde{\hat{\mathbf{F}}}_i\right)\right)^T + (1-\alpha)\gamma_i \gamma_i^T\right)^{-1}$$

5. If $f(\mathbf{x}_i^{(j)}) \geq f\left(\mathbf{x}_{i-1}^{(j)}\right), j=1,...,N,$ then retain $\mathbf{x}_i^{(j)}$ as the updated particle;

   else set $\mathbf{x}_i^{(j)} = \mathbf{x}_{i-1}^{(j)}$

6. If $i < M$, then go to step 3 with $i = i+1$;

   else terminate the algorithm.

It may be noted that, while the prediction step of Eqn. 2.1 appears to be helpful in the exploration, this step may not be practically useful with our scheme, especially as it does not exploit any directional information while exploring. Hence, the global search may be expedited by dropping the prediction step from the ***pseudo-code 1***. Such a modification would typically mean that the number of evaluations of the cost functional would reduce by half. The modified pseudo-code involving no prediction step is provided below.

***The pseudo-code 2:***

1. Follow steps 1 and 2 of pseudo-code 1.

2. *Additive update*

   This step is similar to Step 4 of pseudo-code 1; except that the over-tildes over the variables are no longer needed and $\tilde{\mathbf{x}}_i^{(j)}$ must be replaced by $\mathbf{x}_{i-1}^{(j)}$. However, for clarity, we provide the details of this step below.

   Choose $\alpha \in (0,1)$ and update each particle as:

   $$\mathbf{x}_i^{(j)} = \mathbf{x}_{i-1}^{(j)} + \mathbf{D}_i^{\sigma_2(j)}, \quad j = 1,...,N$$

   where $\mathbf{D}_i^{\sigma_2(j)}$ is the $\sigma_2(j)^{\text{th}}$ correction vector. $\sigma_2(j)$ is as defined in pseudo-code 1.

   $$\mathbf{D}_i^{(j)} := \beta_i \mathbf{G}_i \mathbf{I}_i^{(j)}, \quad j=1,...,N, \quad \mathbf{I}_i^{(j)} := \begin{Bmatrix} \hat{f}_{i-1} - f_{i-1}^{(j)} \\ \mathbf{x}_{i-1}^{\sigma_1(j)} - \mathbf{x}_{i-1}^{(j)} \end{Bmatrix}; \beta_i \text{ and } \sigma_1(j) \text{ as in pseudo-code 1.}$$

$$\mathbf{G}_i := \frac{1}{N}\left\{\left(\mathbf{X}_{i-1} - \hat{\mathbf{X}}_{i-1}\right)\left(\mathbf{F}_{i-1}^T \tau_i - \hat{\mathbf{F}}_{i-1}^T \tau_{i-1} - \Delta\hat{\mathbf{F}}_{i-1}^T \tau_i\right) + \left(\hat{\mathbf{X}}_{i-1}\tau_i - \hat{\mathbf{X}}_{i-1}\tau_{i-1}\right)\left(\mathbf{F}_{i-1} - \hat{\mathbf{F}}_{i-1}\right)^T\right\}$$

$$\left(\alpha \frac{1}{N-1}\left(\left(\hat{\mathbf{F}}_{i-1} - \mathbf{F}_{i-1}\right) - \left(\hat{\mathbf{F}}_{i-1} - \hat{\mathbf{F}}_{i-1}\right)\right)\left(\left(\hat{\mathbf{F}}_{i-1} - \mathbf{F}_{i-1}\right) - \left(\hat{\mathbf{F}}_{i-1} - \hat{\mathbf{F}}_{i-1}\right)\right)^T + (1-\alpha)\gamma_i\gamma_i^T\right)^{-1}$$

3. Same as Step 5 of pseudo-code 1.

4. If $i < M$, go to Step 2 with $i = i + 1$,
   else terminate the algorithm.

## 5. Numerical illustrations

The optimization route to the reconstruction of the system parameters in the chaotic response regimes of nonlinear dynamical systems could be a tricky problem, especially given the locally exponential nature of separation of trajectories starting with closely separated initial conditions within the strange attractor. In order to describe the associated cost functional, consider the following generic form of the dynamical system model:

$$\dot{\mathbf{x}}_t = \mathbf{\Psi}(\mathbf{x}_t, t, \mathbf{\theta}) \tag{5.1}$$

where $\mathbf{x}_t$ is the state space representation of the system response vector, $\mathbf{\Psi}$ is the non-linear vector field and $\mathbf{\theta}$ the unknown parameter vector. Under a time discrete setting $\{t_0, ..., t_M\}$, the objective functional may be defined as:

$$f(\mathbf{\theta}^a) = \sum_{i=0}^{M}\left(\mathbf{x}_i - \mathbf{x}_i^a\right)^T\left(\mathbf{x}_i - \mathbf{x}_i^a\right) \tag{5.2}$$

where $\mathbf{x}_t^a$ is the solution to Eqn. 5.1 when $\mathbf{\theta}$ is replaced by an assumed parameter vector $\mathbf{\theta}^a$. $f(\mathbf{\theta}^a)$ needs to be minimized using a global optimization strategy to arrive at a converged estimate of $\mathbf{\theta}$, which is reported as the unknown parameter vector of the chaotic system. $\mathbf{x}_t$ denotes the 'true' solution (or the 'measurement', even though, unlike stochastic filtering applications, there is currently no noisy component in $\mathbf{x}_t$) obtained using $\boldsymbol{\theta}$. The following two optimization problems have been considered in [21] as test beds for comparing the performance of variants of the PSO. In implementing the proposed optimization method for these problems, we have followed pseudo-code 2.

## 5.1 The Lorenz oscillator

The Lorenz oscillator [22], which is the first known chaotic dynamical system, is a low dimensional model for the convective motion of fluid. The governing differential equation is given as:

$$\dot{\mathbf{x}}_t = \begin{bmatrix} \theta_1(y_t - x_t) \\ \theta_2 x_t - y_t - x_t z_t \\ x_t y_t - \theta_3 z_t \end{bmatrix} \quad (5.3)$$

Here $\mathbf{x}_t := \begin{Bmatrix} x_t \\ y_t \\ z_t \end{Bmatrix}$ is the state vector and $\boldsymbol{\theta} := \begin{Bmatrix} \theta_1 \\ \theta_2 \\ \theta_3 \end{Bmatrix}$ the vector of unknown parameters. While constructing the measurements, the parameter vector is taken as $\boldsymbol{\theta} = \begin{Bmatrix} 10 \\ 28 \\ \frac{8}{3} \end{Bmatrix}$, which corresponds with chaotic behavior. Eqn. 5.3 is numerically solved over $t \in (0, 0.3]$ using $4^{\text{th}}$ order Runge Kutta method with time step $\Delta t = 0.01$. The optimization problem is solved using the proposed methodology and compared with the PSO, a robust and well known random evolutionary optimization scheme available in the literature. To assess the performance of the proposed scheme, initial guesses for the parameters are randomly generated, based on a uniform distribution over $\begin{Bmatrix} -10 \\ -10 \\ 0 \end{Bmatrix} \leq \boldsymbol{\theta}^a = \begin{Bmatrix} \theta_1^a \\ \theta_2^a \\ \theta_3^a \end{Bmatrix} \leq \begin{Bmatrix} 51 \\ 60 \\ 40 \end{Bmatrix}$. For both the methods, the ensemble size is taken as 30. From figures 5.1(a-f), the superior convergence features of the new scheme, aided by a more effectively guided search, is evident. Indeed, the substantively reduced numerical fluctuations via the proposed scheme may be largely attributed to the additive update that, owing to the rigorously encoded directional information, enables a more effective exploration of the search space. While we do not report the details, the error norms in the parameters estimated by our method are consistently lower by several orders vis-a-vis those based on the PSO. In the following figures, the points on the iteration axes at which the parameters are retrieved are indicated by black squares (the associated iteration numbers/parameter values are written within boxes).

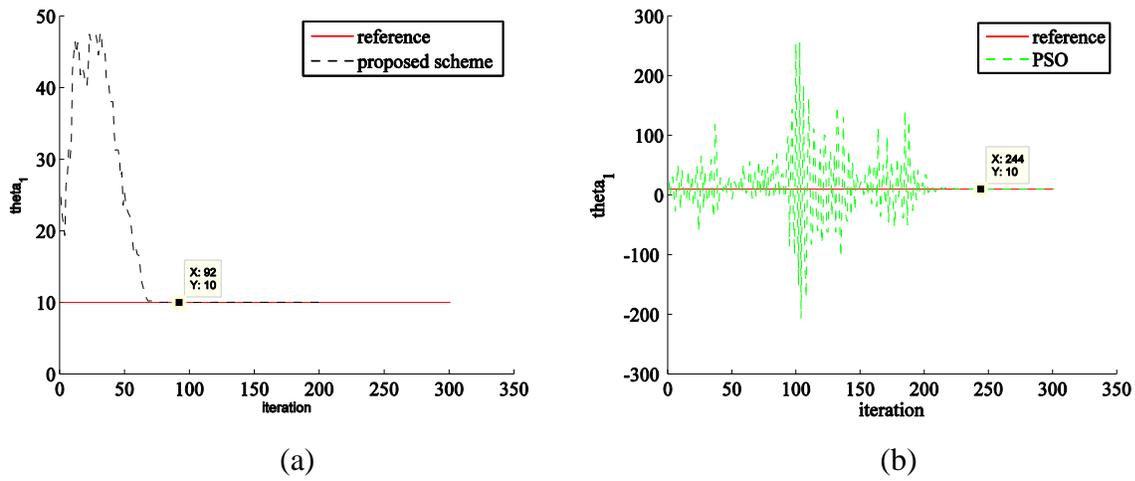

(a)                                             (b)

Fig. 5.1 Evolution of unknown parameter $\theta_1^a$ via (a) the proposed scheme and (b) PSO

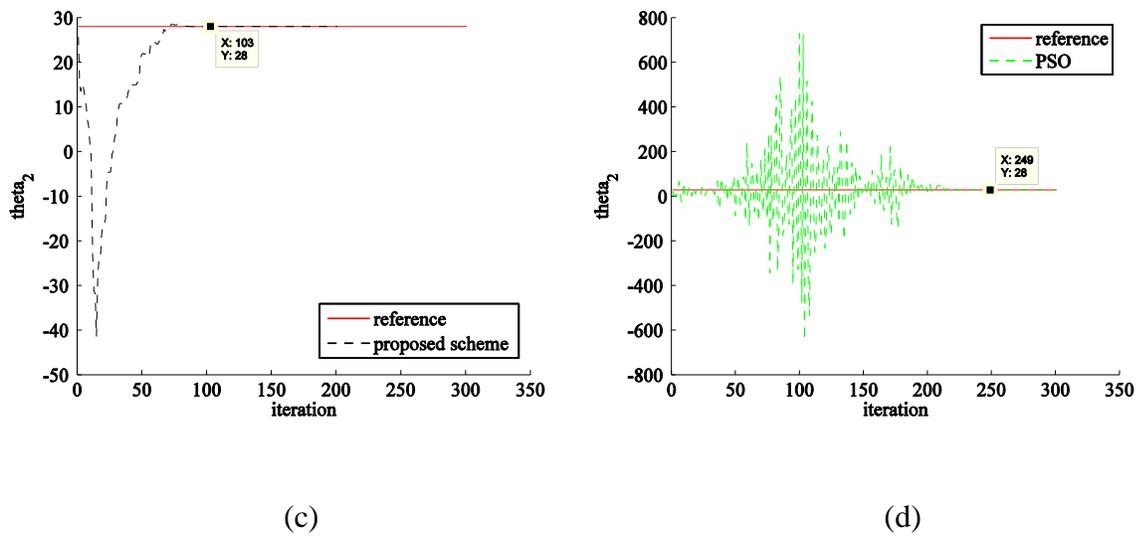

(c)                                             (d)

Fig. 5.1 Evolution of unknown parameter $\theta_2^a$ via (c) the proposed scheme and (d) PSO

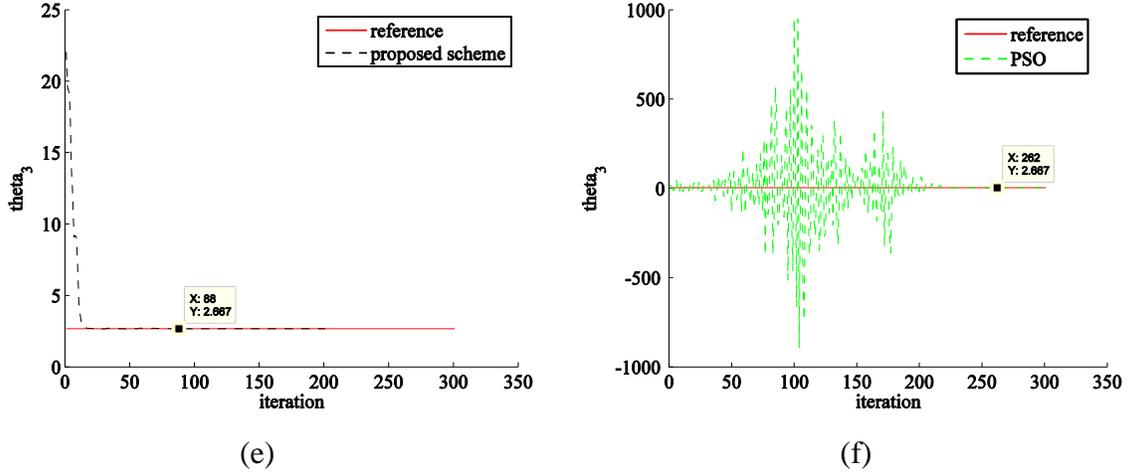

Fig. 5.1 Evolution of unknown parameter $\theta_3^a$ via (e) the proposed scheme and (f) PSO

## 5.2 Chen's oscillator

As another illustration, we consider Chen's oscilator [23, 24] given by:

$$\dot{\mathbf{x}}_t = \begin{bmatrix} \theta_1(y_t - x_t) \\ (\theta_3 - \theta_1)x_t + \theta_3 y_t - x_t z_t \\ x_t y_t - \theta_2 z_t \end{bmatrix} \qquad (5.4)$$

As before, $\mathbf{x}_t := \begin{Bmatrix} x_t \\ y_t \\ z_t \end{Bmatrix}$ is the state vector and $\boldsymbol{\theta} := \begin{Bmatrix} \theta_1 \\ \theta_2 \\ \theta_3 \end{Bmatrix}$ the vector of unknown parameters. Here

$\boldsymbol{\theta} = \begin{Bmatrix} 35 \\ 3 \\ 28 \end{Bmatrix}$ is the reference parameter set that corresponds to Eqn. 5.4 exhibiting chaotic response.

The measurement is generated by numerically integrating this equation with the reference parameters as input. As with the last example, numerical simulations of Eqn. 5.4, required in the global search, are over $t \in (0, 0.3]$ using $4^{\text{th}}$ order Runge-Kutta method with time step $\Delta t = 0.01$. Initial guesses of the parameters are sampled from a uniform distribution over

$$\begin{Bmatrix} -10 \\ -10 \\ 0 \end{Bmatrix} \leq \boldsymbol{\theta}^a = \begin{Bmatrix} \theta_1^a \\ \theta_2^a \\ \theta_3^a \end{Bmatrix} \leq \begin{Bmatrix} 51 \\ 60 \\ 40 \end{Bmatrix}.$$ For both the methods, $N = 30$ is used. The numerical results in figures 5.2(a-f) support similar observations (as in the case of the Lorenz oscillator) regarding the performance of the new scheme and the PSO.

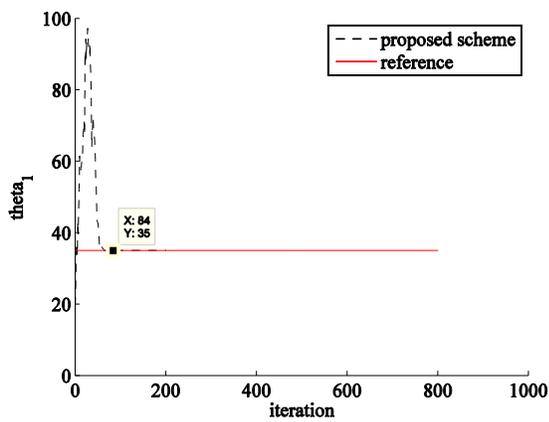
(a)

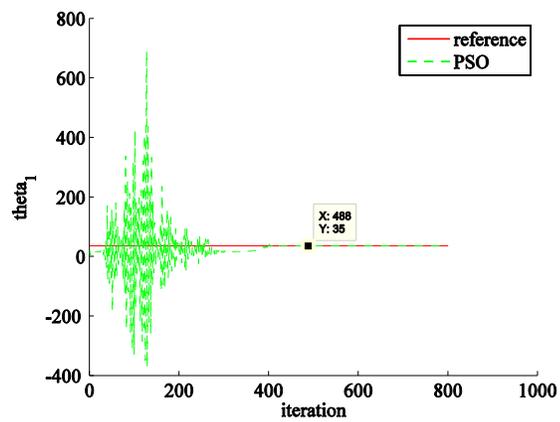
(b)

Fig. 5.2 Evolution of unknown parameter $\theta_1^a$ via (a) the proposed scheme and (b) PSO

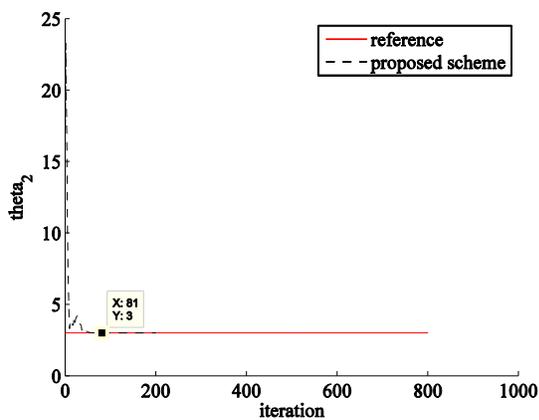
(c)

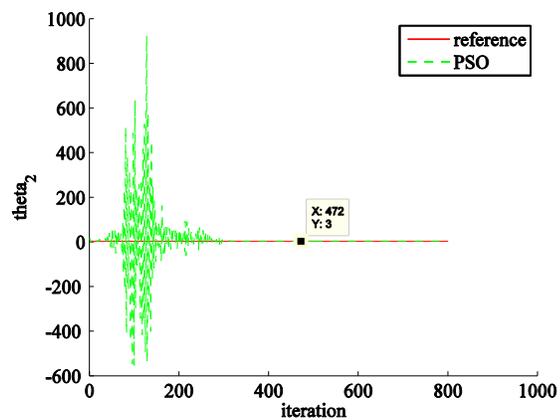
(d)

Fig. 5.2 Evolution of unknown parameter $\theta_2^a$ via (c) the proposed scheme and (d) PSO

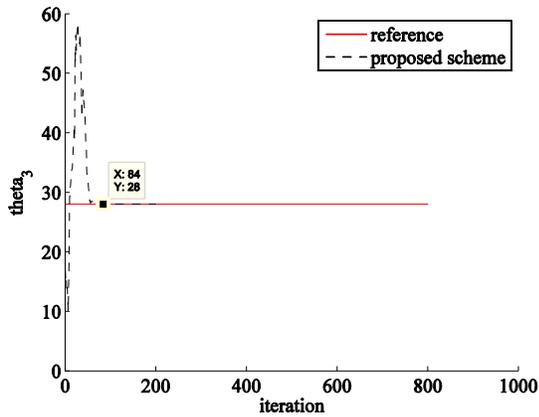
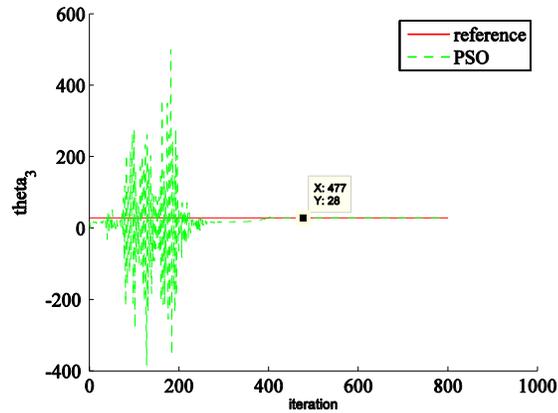

(e)                                              (f)

Fig. 5.2 Evolution of unknown parameter $\theta_3^a$ via (e) the proposed scheme and (f) PSO

## 6. Concluding remarks

Many evolutionary schemes for global optimization, inspired by biological or social observations that may be interesting and elegant in their own right, do not necessarily mark a denouement of rational derivation in approaching the global extremum. This leaves space for questions regarding the performance optimality, especially on the important issue of striking a right balance of convergence speed and exploration efficacy, and the purpose of this article has been to partly address this question through a new proposal for a scheme that has a more formal grounding in the theory of probability and stochastic processes. Specifically, by posing part of the search as a martingale problem, we have derived a derivative-free, additive update procedure that could possibly be interpreted as a non-trivial generalization over the directional derivative based update, classically employed for local extremizations of smooth cost functionals. By way of a further aid to the global search, i.e. in order to preempt possible stalling at local extrema, layers of random perturbation of the martingale problem are also suggested. For most optimization problems, the actual numerical implementation of the method does not however require to employ the random perturbation in its full generality. As demonstrated via a limited set of numerical illustrations that involve minimizing suitable cost functionals *en route* to the parameter recovery of a couple of chaotic oscillators, the proposed method appears to substantively improve upon the performance of one of the most well known evolutionary schemes, the particle swarm optimization.

**Reference**

[1] Holland, J. H. (1975) *Adaptation in natural and artificial systems*, University of Michigan, Ann Arbor, MI, Internal report.

[2] Goldberg, D.E. (1989) *Genetic algorithms in search, optimization and machine learning*, Addison-Wesley, Reading, MA.

[3] Kennedy, J., & Eberhart, R. (1995, November) *Particle swarm optimization, In Proceedings of IEEE international conference on neural networks* 4 (2), 1942-1948.

[4] Dorigo, M., & Birattari, M. (2010) Ant colony optimization, *In Encyclopedia of Machine Learning, Springer US* 36-39.

[5] Fletcher, R. (1987) *Practical Methods of Optimization*, New York, John Wiley.

[6] Chong, E. K. P, Zak, S. H. (2004) *An introduction to optimization*, Singapore, John Wiley & Sons (Asia) Pte. Ltd.

[7] Glover, F; Kochenberger, G.A. (eds.) (2003), *Handbook of Metaheuristics*, Kluwer Academic Publisher.

[8] Van Laarhoven, P. J., & Aarts, E. H. (1987), *Simulated annealing*, 7-15 Springer Netherlands.

[9] Wenzel, W., & Hamacher, K. (1999). Stochastic tunneling approach for global minimization of complex potential energy landscapes, *Physical Review Letters*, 82(15), 3003.

[10] Renders, J.M; Flasse, S. P. (1996) Hybrid methods using genetic algorithms for global optimization, *IEEE Trans. Syst., Man, Cybern. B,* 26 (2) 243–258.

[11] Gordon, N. J; Salmond, D. J; Smith, A. F. M. (1993) Novel approach to nonlinear/non-Gaussian Bayesian state estimation, *Radar and Signal Processing, IEE Proceedings F*, 140, 107-113.

[12] Arulampalam, S; Maskell, N; Gordon, N; Clapp, T. (2002) A tutorial on particle filters for online nonlinear/non-Gaussian Bayesian tracking, *IEEE Transactions on Signal Processing*, 50, 174-188.

[13] Snyder, C., Bengtsson, T., Bickel, P., & Anderson, J. (2008) Obstacles to high-dimensional particle filtering, *Monthly Weather Review*, 136(12).

[14] Oksendal, B.K. (2003) *Stochastic Differential Equations-An Introduction With Applications*, 6th ed., Springer, New York.

[15] Stroock, D. W., & Varadhan, S. R. (1972) On the support of diffusion processes with applications to the strong maximum principle, *In Proceedings of the Sixth Berkeley Symposium on Mathematical Statistics and Probability (Univ. California, Berkeley, Calif.,* 1970/1971) 3, 333-359.

[16] Freidlin, M. I., Szücs, J., & Wentzell, A. D. (2012). *Random perturbations of dynamical systems* 260, Springer.

[17] F. C. Klebaner (2001), *Introduction to stochastic calculus with applications*, London, UK: Imperial College Press.

[18] Kushner, (1964) H. J. On the differential equations satisfied by conditional probablitity densities of Markov processes, with applications. *SIAM J. Series A: Control* 2.1: 106-119.

[19] Sarkar, S., Chowdhury, S. R., Venugopal, M., Vasu, R. M., & Roy, D. (2014). A Kushner–Stratonovich Monte Carlo filter applied to nonlinear dynamical system identification, *Physica D: Nonlinear Phenomena*, 270, 46-59.

[20] Sarkar, S., & Roy, D. (2014) An Ensemble Kushner-Stratonovich (EnKS) Nonlinear Filter: Additive Particle Updates in Non-Iterative and Iterative Forms. *arXiv preprint arXiv*:1402.1253.

[21] Sun, J., Zhao, J., Wu, X., Fang, W., Cai, Y., & Xu, W. (2010) Parameter estimation for chaotic systems with a Drift Particle Swarm Optimization method, *Physics Letters A*, 374(28), 2816-2822.

[22] Lorenz, E.N. (1963) Deterministic nonperiodic flow, Journal of Atmospheric Sciences, 20(2), 130-141.

[23] Chen. G. & Ueta, T. (1999) Yet another chaotic attractor, *International Journal of Bifurcation and Chaos*, 9(07), 1465-1466.

[24] Lu, J. & Chen, G. (2002) A new chaotic attractor coined, *International journal of Bifurcation and Chaos*, 12(03), 659-661.

## Appendix I: Derivation of the extremal equation

The gain based update equation for a bounded, at least twice differentiable function $\phi_\tau := \phi(\mathbf{x}_\tau)$ of $\mathbf{x}_\tau$ may be arrived at by expanding $\phi(\mathbf{x}_\tau)\Lambda_\tau$, where $\tau \in (\tau_{i-1}, \tau_i]$, using Ito's formula (the stochastic counterpart of Taylor's expansion):

$$d[\phi_\tau \Lambda_\tau] = \phi_\tau d\Lambda_\tau + d\phi_\tau \Lambda_\tau + \langle d\phi_\tau, d\Lambda_\tau \rangle \tag{A1}$$

$\langle \cdot \rangle$ denotes the quadratic covariation. A further expansion leads to:

$$d[\phi_\tau \Lambda_\tau] = \phi_\tau \Lambda_\tau h_\tau d\breve{f}_\tau + \Lambda_\tau \phi_\tau'^T d\mathbf{x}_\tau + \frac{1}{2} \Lambda_\tau \langle d\mathbf{x}_\tau, \phi_\tau'' d\mathbf{x}_\tau \rangle \tag{A2}$$

By explicitly writing the term $\langle d\mathbf{x}_\tau, \phi_\tau'' d\mathbf{x}_\tau \rangle$ we get:

$$d[\phi_\tau \Lambda_\tau] = \phi_\tau \Lambda_\tau h_\tau d\breve{f}_\tau + \Lambda_\tau \phi_\tau'^T d\mathbf{x}_\tau + \frac{1}{2} \sum_{j,k=1}^n \sum_{l=1}^d \left( \frac{\partial^2 \phi}{\partial x^j \partial x^k} \right)_\tau g^{jl} g^{kl} d\tau \tag{A3}$$

The incremental form as in Eqn. A3 may be given the following integral representation:

$$\phi_\tau \Lambda_\tau = \phi_{i-1} \Lambda_{i-1} + \int_{\tau_{i-1}}^\tau \Lambda_s \phi_s h_s d\breve{f}_s + \int_{\tau_{i-1}}^\tau \Lambda_s \left( \phi_s' d\xi_s + \frac{1}{2} \sum_{j,k=1}^n \sum_{l=1}^d \left( \frac{\partial^2 \phi}{\partial x^j \partial x^k} \right)_s g^{jl} g^{kl} \right) ds \tag{A3}$$

In deriving Eqn. A3, Eqn. 2.1 is made use of. Taking conditional expectation with respect to $\mathcal{N}_\tau$ under *Q* we get:

$$E_Q\left[\phi_\tau \Lambda_\tau \mid \mathcal{N}_\tau\right] = E_Q\left[\phi_{i-1}\Lambda_{i-1} \mid \mathcal{N}_\tau\right] + E_Q\left[\left\{\int_{\tau_{i-1}}^{\tau} \Lambda_s \phi_s h_s d\breve{f}_s\right\} \mid \mathcal{N}_\tau\right] + E_Q\left[\left\{\int_{\tau_{i-1}}^{\tau} \Lambda_s \phi'_s d\xi_s\right\} \mid \mathcal{N}_\tau\right]$$

$$+ \frac{1}{2} E_Q\left[\left\{\int_{\tau_{i-1}}^{\tau} \frac{1}{2} \sum_{j,k=1}^{n} \sum_{l=1}^{d} \left(\frac{\partial^2 \phi}{\partial x^j \partial x^k}\right)_s g^{jl} g^{kl} ds\right\} \mid \mathcal{N}_\tau\right]$$

(A4)

Using Fubini's theorem:

$$E_Q\left[\phi_\tau \Lambda_\tau \mid \mathcal{N}_\tau\right] = E_Q\left[\phi_{i-1}\Lambda_{i-1} \mid \mathcal{N}_\tau\right] + \int_{\tau_{i-1}}^{\tau} E_Q\left[\Lambda_s \phi_s h_s \mid \mathcal{N}_s\right] d\breve{f}_s + \int_{\tau_{i-1}}^{\tau} E_Q\left[\Lambda_s \phi'_s \mid \mathcal{N}_s\right] d\xi_s$$

$$+ \frac{1}{2}\int_{\tau_{i-1}}^{\tau} E_Q\left[\sum_{j,k=1}^{n}\sum_{l=1}^{d}\left(\frac{\partial^2 \phi}{\partial x^j \partial x^k}\right)_s g^{jl}g^{kl} \mid \mathcal{N}_s\right] ds$$

(A5)

Noting that $\int_{\tau_{i-1}}^{\tau} E_Q\left[\{\Lambda_s \phi'_s\} \mid \mathcal{N}_s\right] d\xi_s = 0$ and for notational convenience denoting the un-normalized conditional expectation operator, $E_Q\left[(\cdot)_\tau \Lambda_\tau \mid \mathcal{N}_s\right]$ as $\sigma_\tau(\cdot)$ we arrive at the following equation:

$$\sigma_\tau(\phi) = \sigma_{i-1}(\phi) + \int_{\tau_{i-1}}^{\tau} \sigma_s(\phi h) d\breve{f}_s + \frac{1}{2}\int_{\tau_{i-1}}^{\tau} \sigma_s\left(\sum_{j,k=1}^{n}\sum_{l=1}^{d}\left(\frac{\partial^2 \phi}{\partial x^j \partial x^k}\right)_s g^{jl}g^{kl}\right) ds$$

(A6)

Eqn. A6 finds its equivalence with the Zakai equation well known in stochastic filtering. An incremental representation of Eqn. A6 may be given as:

$$d\sigma_\tau(\phi) = \sigma_\tau(\phi h) d\breve{f}_\tau + \frac{1}{2}\sigma_\tau\left(\sum_{j,k=1}^{n}\sum_{l=1}^{d}\left(\frac{\partial^2 \phi}{\partial x^j \partial x^k}\right) g^{jl}g^{kl}\right) d\tau$$

(A7)

In order to obtain the normalized conditional law, i.e. $\pi_\tau(\phi) = \frac{\sigma_\tau(\phi)}{\sigma_\tau(1)}$, it is expanded using Ito's formula as given below:

$$d\pi_\tau(\phi) = \frac{d\sigma_\tau(\phi)}{\sigma_\tau(1)} + \sigma_\tau(\phi) d\left(\frac{1}{\sigma_\tau(1)}\right) + \left\langle d\sigma_\tau(\phi), d\left(\frac{1}{\sigma_\tau(1)}\right)\right\rangle$$

(A8)

$d\left(\dfrac{1}{\sigma_\tau(1)}\right)$ may be expanded as:

$$d\left(\dfrac{1}{\sigma_\tau(1)}\right) = -\dfrac{1}{\sigma_\tau^2(1)} d\sigma_\tau(1) + \dfrac{1}{\sigma_\tau^3(1)} \langle d\sigma_\tau(1), d\sigma_\tau(1) \rangle \tag{A9}$$

Putting $\phi = 1$ in Eqn. A7, we get an Ito expansion for $d\sigma_\tau(1)$, which is given below:

$$d\sigma_\tau(1) = \sigma_\tau(h) d\breve{f}_\tau \tag{A10}$$

Using Eqn. A10 in Eqn. A9:

$$d\left(\dfrac{1}{\sigma_\tau(1)}\right) = -\dfrac{\pi_\tau(h)}{\sigma_\tau(1)} d\tilde{f}_\tau + \dfrac{\pi_\tau^2(h)}{\sigma_\tau(1)} d\tau \tag{A11}$$

Using Eqn. A7 and A11 in Eqn. A8 we get:

$$d\pi_\tau(\phi) = \pi_\tau(\phi h) d\breve{f}_\tau + \dfrac{1}{2}\pi_\tau\left(\sum_{j,k=1}^{n}\sum_{l=1}^{d}\left(\dfrac{\partial^2 \phi}{\partial x^j \partial x^k}\right) g^{jl} g^{kl}\right) d\tau$$
$$+ \left(-\pi_\tau(\phi)\pi_\tau(h) d\breve{f}_\tau + \pi_\tau(\phi)\pi_\tau^2(h) d\tau\right) - \pi_\tau(\phi h)\pi_\tau(h) d\tau$$

$$\tag{A12}$$

Thus we arrive at the integro-differential equation describing the evolution of the normalized conditional estimate $\pi_\tau(\phi)$ given as:

$$d\pi_\tau(\phi) = \dfrac{1}{2}\pi_\tau\left(\sum_{j,k=1}^{n}\sum_{l=1}^{d}\left(\dfrac{\partial^2 \phi}{\partial x^j \partial x^k}\right) g^{jl} g^{kl}\right) d\tau + \left(\pi_\tau(\phi h) - \pi_\tau(\phi)\pi_\tau(h)\right)\left(d\breve{f}_\tau - \pi_\tau(h) d\tau\right) \tag{A13}$$

Eqn. A13 has its parallel in the Kushner-Stratonovich equation, which is well known in stochastic filtering. Since we are typically interested in the evolution of the conditional estimate of $\mathbf{x}_\tau$, i.e. $\phi$ is the identity function, Eqn. A13 may be simplified as:

$$d\pi_\tau(\mathbf{x}) = \left(\pi_\tau(\mathbf{x}h) - \pi_\tau(\mathbf{x})\pi_\tau(h)\right)\left(d\breve{f}_\tau - \pi_\tau(h)d\tau\right) \tag{A14}$$

Replacing $h(\mathbf{x}_\tau)$ by $\rho_\tau^{-1}f(\mathbf{x}_\tau)$ in Eqn. A14, we get:

$$d\pi_\tau(\mathbf{x}) = \left(\pi_\tau(\mathbf{x}f) - \pi_\tau(\mathbf{x})\pi_\tau(f)\right)\left(\rho_\tau\rho_\tau^T\right)^{-1}\left(d\breve{f}_\tau - \pi_\tau(f)d\tau\right) \tag{A15}$$